\documentclass{appolb}
\usepackage{epsfig}

\begin{document}
\title{CAN COSMIC RAYS PROVIDE SIGN OF STRANGELETS?
\thanks{Invited talk presented by Z. W\l odarczyk (wlod@pu.kielce.pl)
at the XXVII Mazurian Lakes School of Physics {\it Growth Points of
Nuclear Physics A.D. 2001}, September 2-9, 2001, Krzy\.ze, Poland.} 
}
\author{M. Rybczy\'nski, Z. W\l odarczyk
\address{Institute of Physics, \'Swi\c{e}tokrzyska Academy, Kielce, Poland}
\and G. Wilk
\address{The Andrzej So\l tan Institute of Nuclear Studies, Warsaw, Poland}
}
\maketitle
\begin{abstract}
We discuss the possible imprints of Strange Quark Matter (SQM) in
cosmic ray data. In particular, we investigate the propagation of SQM
through the atmosphere and discuss: i) direct candidates for
strangelets, ii) exotic events interpreted as signals of SQM and iii)
muon bundles and delayed neutrons in Extensive Air Showers. The
physics and astrophysics of SQM is shortly reviewed. We point out the
possibility that extreme energy cosmic rays are the results of the
decay of unstable primordial objects. Finally, the abundance of
possible candidates for strangelets and their mass spectrum are
estimated and compared with the astrophysical limits and prospects of
the possible observation of SQM in accelerator experiments are
outlined.
\end{abstract}

\section{Introduction}
In the astrophysical literature \cite{Klin} one can find a number of
phenomena which can be regarded as a possible manifestation of the
existence of the so called Strange Quark Matter (SQM) \cite{Wit}
(existing in the form of lumps called strangelets). This is extremely
interesting possibility of apparently new stable form of matter (it
can decay only via weak interactions, which for a bulk of matter
consisting strangelet are very inefficient in reduction its size). In
particular one observes (cf. \cite{WWj} for relevant references):  
\begin{itemize}
\item anomalous cosmic ray bursts from {\it Cygnus X-3},
\item extraordinary high luminosity gamma ray burst from the {\it
supernova remnant N49\/} in the Large Magellanic Cloud,
\item or the so called {\it Centauro\/} events, which are characterized
by anomalous composition of secondary particles with almost no neutral
pions present.
\end{itemize}
There were also attempts to find lumps of SQM, called {\it
strangelets,\/} in terrestrial experiments devoted to search for the
Quark Gluon Plasma (QGP) state of matter but so far without apparent
success (cf. Ref. \cite{Klin} and last Section below). This fact,
however, does not preclude sensibility  of searching for strangelets
in cosmic ray experiments, which deal with strangelets formed in
completely different astrophysical mechanisms \cite{Wit}, besides one
witnesses their production proceeding  in collisions of the original
CR flux with atmospheric nuclei \cite{Pan}. However, any SQM produced
at very early stage of the history of the Universe would have
evaporated long time ago due to the action of weak interactions
\cite{Alc}. On the other hand SQM is probably continuously produced in
neutron stars with a super-dense quark surface and in quark stars
with a thin nucleon envelope \cite{Wit}. Collisions of such objects
could therefore produce small lumps of SQM, strangelets with $10^{2}
< A < 10^{6}$, permeating the Galaxy and possibly reaching also
Earth, i.e., {\it a priori\/} being detectable here. In this
presentation we demonstrate how strangelets (not much different in
size from the similar lumps of normal nuclear matter) can still
penetrate deeply in the atmosphere. We estimate the initial flux of
strangelets entering the atmosphere, and finally we point out the
possibility that extreme energy cosmic rays are the results of the
decay of unstable primordial objects which can possibly be
identified with strangelets. We shall list also the presently
running and planned experiments looking for SQM being produced
already at accelerators and summarize results obtained so far. 

\section{Some features of strangelets}
Typical SQM consists of roughly equal number of up {\it (u),\/} down
{\it (d)\/} and strange {\it (s)\/} quarks, and it has been found to
be the true ground state of QCD \cite{Alc}, i.e., it is absolutely
stable at high mass numbers $A$ and, because the energy per baryon in
SQM could be smaller than that in ordinary nuclear matter, it would
be more stable than the most tightly bound $^{56}Fe$ nucleus. The
measure of stability of strangelets is provided by the so called
separation energy $dE/dA$, i.e., energy which is required to remove a
single baryon from a strangelet. There exists some critical size
given by a critical value of $A = A_{crit}$ (vary from $A_{crit} =
300$ to $400$ depending on the various choices of parameters
\cite{Alc}) such that for $A > A_{crit}$ strangelets are absolutely
stable against neutron emission \cite{Schaf}. Below this limit
strangelets decays rapidly by evaporating neutrons.\\  

The small value of the charge to mass ratio, $Z/A \sim A^{1/3}$,
expected in the case of strangelets \cite{Alc,Kas}, provide us with
main criterion for their discrimination among other debris when
searching cosmic rays for such nuclearities.\\ 

As we have demostrated in \cite{WWj} the rescaled radius $r_0$ of
strangelets (which follow $A$-dependence typical for nuclei, i.e.,
$R=r_0 A^{1/3}$) is comparable to this of ordinary nuclei.
Considering a lump of SQM visualized after \cite{Alc} as Fermi gas of
{\it u, d\/} and {\it s\/} quarks, with total mass number $A$ which
is  confined in a spherical volume $V \propto A$, the rescaled radius
$r_{0}$ is determined by the number density of strange matter
\cite{Alc}. For the values commonly accepted for SQM  (like the mass
of the strange quark $m \cong 150$ $MeV$ and the chemical potential
$\mu \cong 300$ $MeV$), the values of $r_{0}$ of the strangelets are
comparable with that for the ordinary nuclear matter \cite{WWj}
(being only a bit smaller, with differences not exceeding 10\% -
20\%). However, because the mass number $A$ of strangelets is much
larger than the mass number of ordinary nuclei, their expected
geometrical cross sections are also much larger than those for
normal nuclei.

\section{Propagation of Strangelets in the Atmosphere}

It does not mean, however, that some form of SQM does not penetrate
deep in the atmosphere to be finally registered. The apparent
contradiction between its "normal" size and strong penetrability can
be resolved in very simple manner. Strangelets reaching so deeply
into atmosphere are formed in many successive interactions with air
nuclei following one of the proposed possible scenarios:
\begin{itemize}
\item[$(i)$] either an initially small strangelet picks up mass from
the collisions with air atoms which during its passage through Earth
atmosphere  \cite{Raha}, 
\item[$(ii)$] or an initially very heavy lump of SQM entering our
atmosphere (with $A$ of the order of $10^3$) decreases due to
subsequent collisions with air nuclei, until its mass reaches a
critical value $A_{crit}$  at which point it disintegrates
\cite{WWj,WWh}. 
\end{itemize}
In what follows we explore the second scenario developed by us
recently. In this way one can accomodate both the most probably
"normal" mean free paths for successive interactions and final large
penetrating depth. Such scenario is fully consistent with all present
experiments \cite{WWh}. The scenario proposed and tested in
\cite{WWj} was that in the interaction of strangelet with target
nucleus all quarks of the target (which are located in the
geometrical intersection of the two colliding nuclei) are involved.
It is assumed that each quark from the target interacts with only one
quark from the strangelet; i.e., during interaction the mass number
of strangelet is diminished to the value equal to $A - A_{t}$ at
most. This procedure continuous unless either strangelet reaches
Earth or (most probably) disintegrates at some depth $h$ of the
atmosphere reaching $A(h) = A_{crit}$. This result, in a first
approximation (in which $A_{t} \ll A_{crit} < A_{0}$), in the total
penetration depth of order of $\Lambda \cong
\frac{4}{3}\lambda_{NA_{t}}(A_{0}/A_{t})^{1/3}$. Fig. 1 shows at what
depth strangelets start to become critical whereas Fig. 2 exposes the
expected number of nucleons released from strangelet at depth $h$ of
the atmosphere.

\vspace{-1.cm}
\begin{figure}[h]
\begin{minipage}[t]{0.475\linewidth}
\centering
\includegraphics[height=7.cm,width=7.cm]{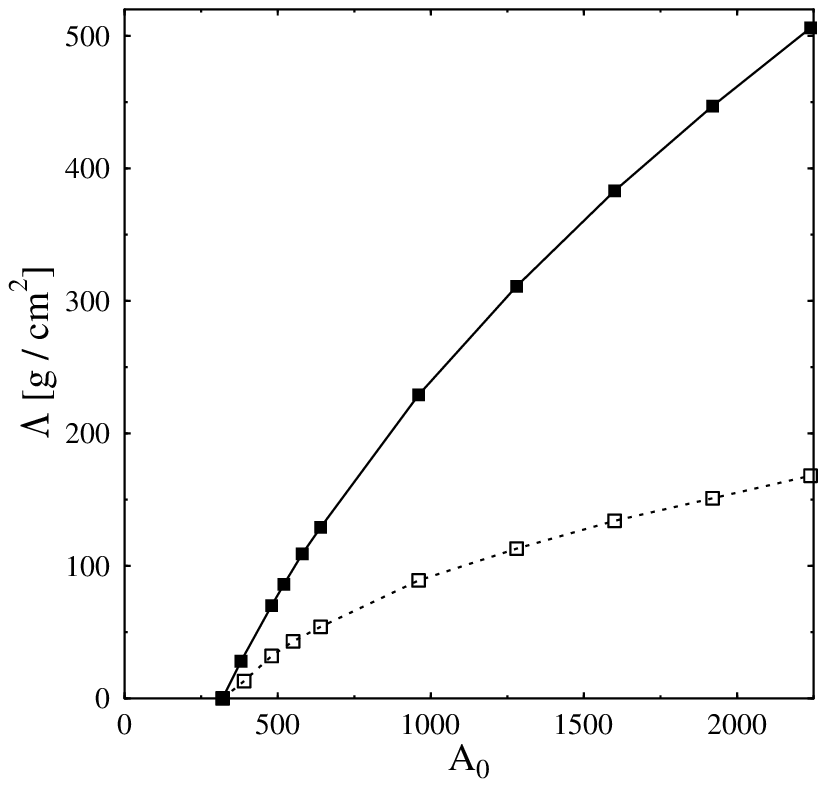}
\caption{Atmospheric length $\Lambda$ after which initial strangelets 
reaches its critical size $A_{crit}=320$ drawn as a function of 
its initial mass number $A_0$ for SM (solid) and TM (dashed) models
of interaction with air nuclei. Consecutive squares indicate points
where $A_0/A_{crit}= 2,3,\dots$, (for $A_0>600$).}
\end{minipage}\hspace{2mm}
\begin{minipage}[t]{0.475\linewidth}
\centering
\includegraphics[height=7.cm,width=7.cm]{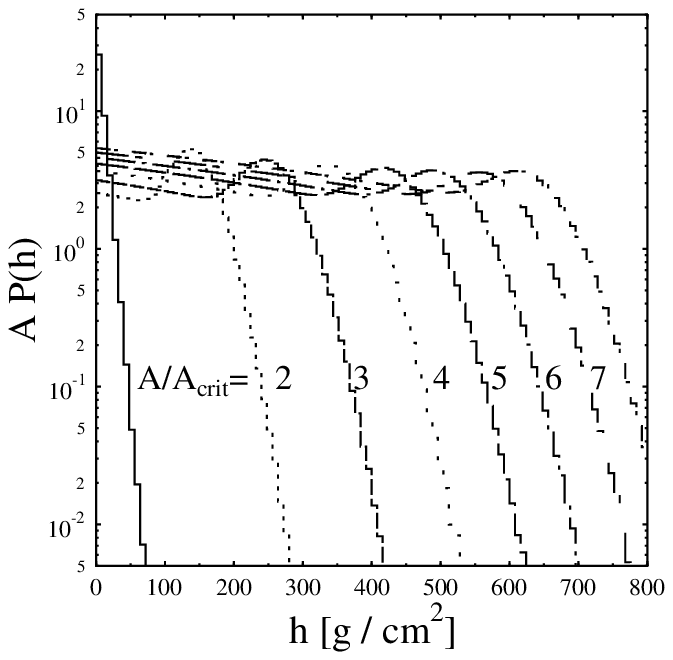}
\caption{Number of nucleons relased in $1 \: g/cm^2$ at depth $h$
of the atmosphere from the strangelet with mass number ratios
$A_0/A_{crit}=1,2,...,8$, respectively (in SM model).}
\end{minipage}
\end{figure}
In numerical estimations provided in \cite{WWj,WWh} in addition to
the above "standard model" (SM) of collisions of strangelets with the
air nuclei we have also considered the so called "tube-like model"
(TM) in which all quarks from geometrical intersection region, both
from projectile and target, participate in the collision. However,
this variant, which leads to the maximal possible destruction of
quarks in strangelets, i.e., to their maximal decreasing, can not
describe adequately experimental data (cf. Fig. 3).

\section{Cosmic nuclearities}

There are several reports suggesting direct candidates for SQM. In
particular, the following anomalous massive particles, which can be
interpreted  as strangelets, have been apparently observed in cosmic
ray (CR) experiments:

\vspace{-1.cm}
\begin{figure}[h]
\begin{minipage}[t]{0.475\linewidth}
\centering
\includegraphics[height=7.cm,width=7.cm]{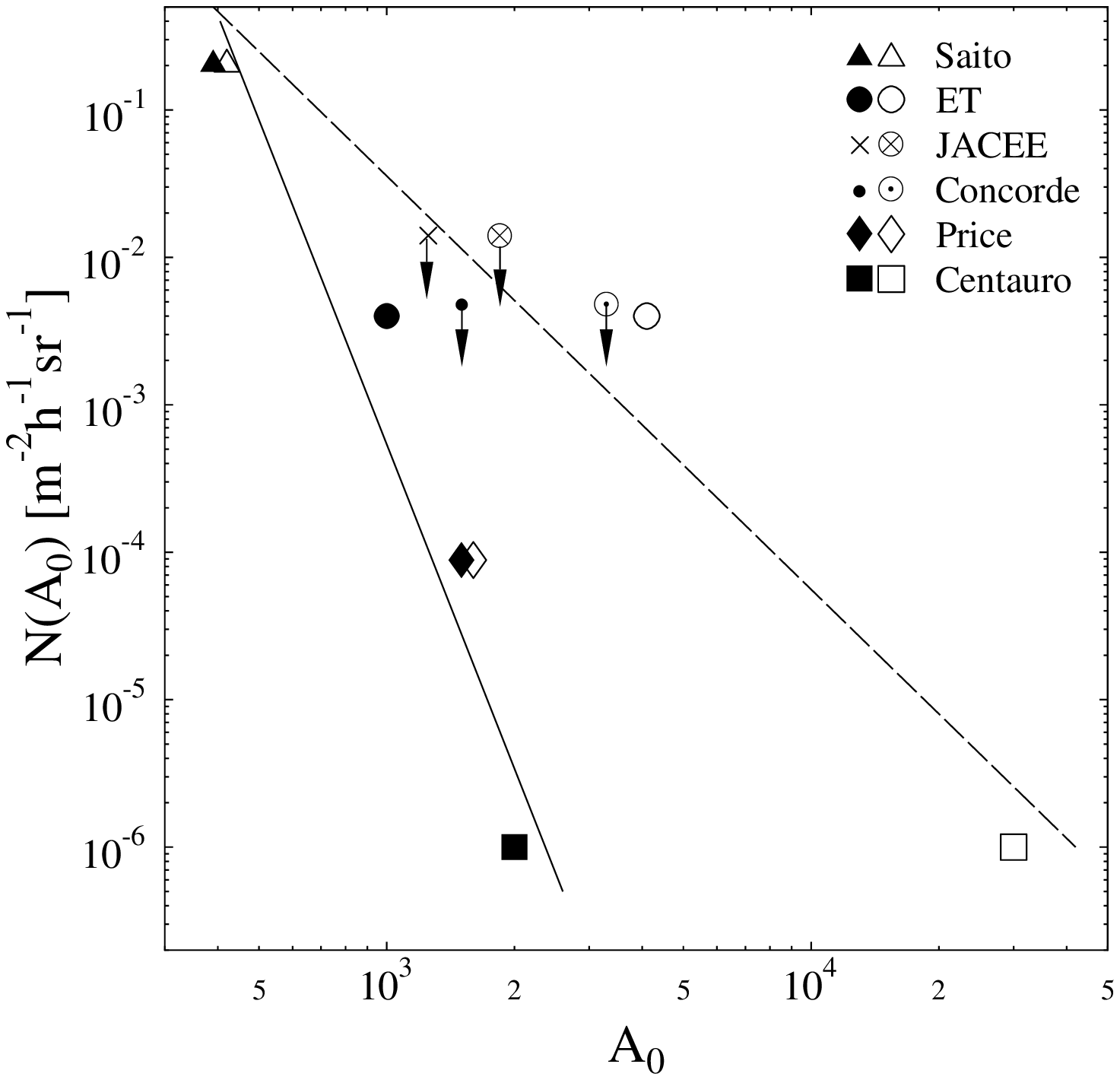}
\caption{The estimation of the expected flux of strangelets on the
border of atmosphere, $N(A_{0})$, as a function of their mass number as
obtained from SM (full symbols; solid line for power fits) and TM
(empty symbols; dashed line). See \cite{WWj,WWh} for further details.}
\end{minipage}\hspace{2mm}
\begin{minipage}[t]{0.475\linewidth}
\centering
\includegraphics[height=7.cm,width=7.cm]{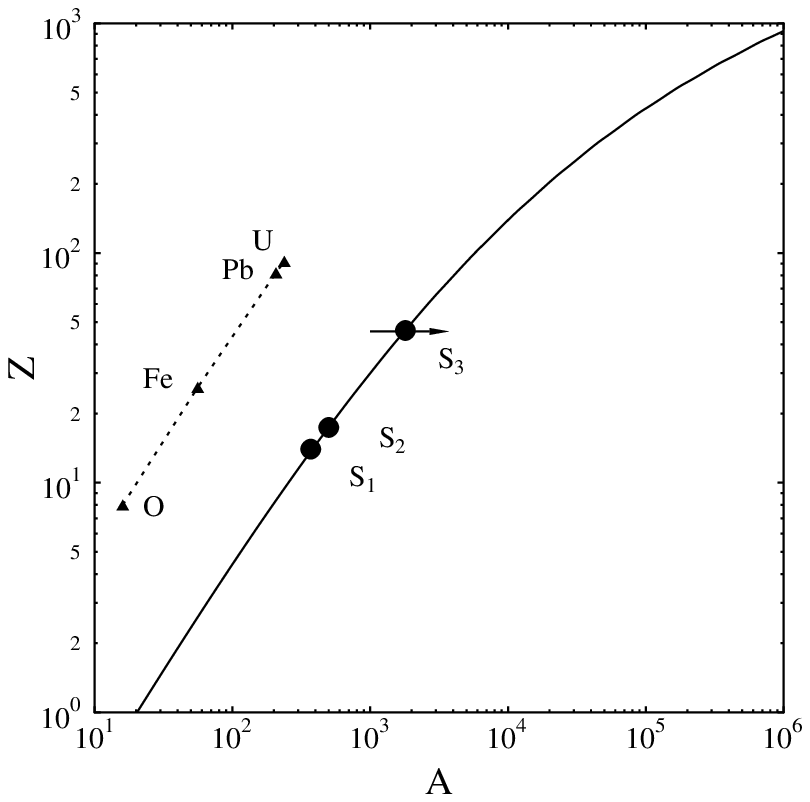}
\caption{The relation of $Z$ and $A$ for SQM (solid line) in
comparison with the same for normal nuclei (dashed line). 
Notice that candidates for strangelets ($S1$ \cite{Sai}, $S2$
\cite{Ichi} and $S3$ \cite{Pri}) lie on theoretical line.}
\end{minipage}
\end{figure}
\begin{itemize}
\item[$(i)$] In counter experiment devoted to study primary CR nuclei two
anomalous events have been observed ({\it Saito}) \cite{Sai} with
values of charge $Z \cong 14$ and of mass numbers $A \cong 350$ and
$A \cong 450$ respectively.
\item[$(ii)$] The so called Price's event \cite{Pri} with $Z \cong 46$ and $A >
1000$, regarded previously as possible candidate for magnetic monopole.
\item[$(iii)$
] The so called Exotic Track event ({\it ET}) with $Z \cong 20$ and
$A \cong 460$ has been reported in \cite{Ichi}. The name comes from
the fact that the projectile causing that event has apparently
traversed $\approx 200$ g/cm$^{2}$ of atmosphere.
\end{itemize}
It is remarkable that all possible candidates for SQM have mass numbers
near or slightly exceeding $A_{crit}$ (it is also argued that {\it
Centauro\/} \cite{Cent} event, regarded to be possible candidate for
strangelet, contains probably $\approx 200$ baryons \cite{Bjo}). Also
the values of $Z$ and $A$ mentioned above are fully consistent with
the existing theoretical estimations for $Z/A$ ratio, which is
characteristic for the SQM \cite{Kasu}, cf. Fig. 4. 

Using our scenario of strangelet propagation \cite{WWj,WWh} and
experimental data taken at different atmospheric depths we can
estimate the flux of strangelets reaching our atmosphere, c.f. Fig.
3. The experimental data {\it Saito, ET\/} \cite{Sai,Ichi} and {\it
Centauro\/} \cite{Cent} on measured fluxes on different atmospheric
depths as well as the corresponding upper limits (no strangelets
found so far), {\it JACEE\/} \cite{JACEE} and {\it Concorde,\/}
\cite{Conc} are processed. Notice, that Price's data favor standard
model. In terms of fits (for 3 points: {\it Saito, ET, Centauro\/})
one gets: $\sim A_{0}^{-7.5}$ for SM and $\sim A_{0}^{-2.8}$ for TM
(dashed line). The choice of the power-like form of standard model was
dictated by the analogy to nuclear fragmentation and the expectation
that decay (fragmentation) of a strange star after its collision will
result in the production of strangelets with similar distribution of
mass. 

The analysis \cite{WWj,WWh} of the above listed candidates for SQM
indicate that the abundance of strangelets in primary cosmic ray flux
is roughly $F_{S}(A_{0} = A_{crit})/F_{tot} = 2.4 \cdot 10^{-5}$ at
the same energy per particle.

\section{Exotic events}

We would like to discuss now shortly a number of existing
experimental results obtained by Emulsion Chamber experiments, which
can be regarded as exotic (i.e., not encountered so far in accelerator
experiments and therefore still waiting for its proper
understanding) \cite{Cal}. Those are, for example, centauro species,
superfamilies with "halo", the strongly penetrating component, etc..
As already mentioned, they can hardly be explained in terms of
standard ideas about hadronic interactions, which we have learned so
far from accelerator experiments. Our Monte Carlo simulations show,
however, that these phenomena could not originate from any kind of
statistical fluctuations of "normal" hadronic interactions,
indicating therefore that either some new mechanism of interaction or
new primaries might appear in the high energy interactions. Our
attitude is to attribute all those effects to  action of strangelets.

\subsection{Mini-cluster}
The transition curves of anomalous cascades exhibit surprising
features: a strong penetrating nature associated with very slow
attenuation and appearance of many maxima with small distances between
them (about 2-3 times smaller than the calculated distances for the
"normal" hadron cascades). We investigate \cite{Gla} the possible
connection between this extremely penetrating component, frequently
accompanying the cosmic ray exotic phenomena, and the hypothesis of
the formation of strangelets taking place in the process of
strangeness distillation, being the last stage of the evolution of
the quark matter droplet. We find that many-maxima long-range
cascades observed in the homogeneous lead emulsion chamber
experiments can be produced in the process of strangelet penetration
through the apparatus. The bundle of hadrons provides the possible
explanation for the anomalous (strongly penetrating) showers. In
order to explain the mutual distance distribution of the sub-showers
maxima positions, we need only a few $(\sim 7)$ hadrons concentrated in
the very forward region. The assumption of strangelet with $A = 15$
penetrating through the chamber and evaporate neutrons leads to the
formation of the long-range many-maxima cascades similar to those
observed in the experiment. 

The existing experimental data, however, are not sufficient to decide
if they are produced by (stable or unstable) strangelets. The CASTOR
detecting system, proposed as a subsystem of the ALICE experiment at
LHC could help in solving the existing uncertainties \cite{Ang}.

\subsection{Centauros}
The Centauro and mini-Centauro events, characterized by the extreme
imbalance between hadronic and gamma-ray components among the produced
secondaries, are the best known examples of numerous unusual events
reported in cosmic-ray experiments. There are many attempts to explain
them (different types of isospin fluctuations or formation of
disoriented chiral condensate, multiparticle Bose-Einstein
correlations, strange quark matter formation or interaction). It was
shown that families recorded at mountain altitudes are insensitive to
any isospin fluctuations. Centauro-like phenomena require deeply
penetrating component in cosmic rays. We claim that they can be
products of strangelets penetrating deeply into atmosphere and
evaporating neutrons \cite{WWn}. Both the flux ratio of Centauros
registered at different depths and the energy distribution within
them can be successfully described by such concept.

\subsection{Coplanar emission}
Phenomenon of alignment of structural objects of gamma-hadron families
near a stright line in the plane at the target diagram was first observed
during examination of multicore halos and, later, when observing
distinguished energetic cores (i.e., halos, energetic hadrons, high
energy gamma quanta or narrow particle groups) \cite{Bor}. The excess
of aligned families found in these cases exceeds any known conventional
concept of interaction. Many attempts to interpret this phenomenon of
coplanar emission were undertaken. However, up to now no satisfactory
explanation exists. The trouble is that this alignment occurs in
spite of the fact that there is always substantial number of
interactions contributing to family formation. The long-flying
component with mean free path of the few hundreds of $g/cm^{2}$ is
required. As a tentative explanation the arrival of strangelets with
high spin $(J \sim A^{2})$ and gradual dispersion of mass $A(h)$
along their way through the atmosphere can be given.

\section{Families and EAS induced by strangelets}

\vspace{-1cm}
\begin{figure}[h]
\begin{minipage}[t]{0.475\linewidth}
\centering
\includegraphics[height=7.cm,width=7.cm]{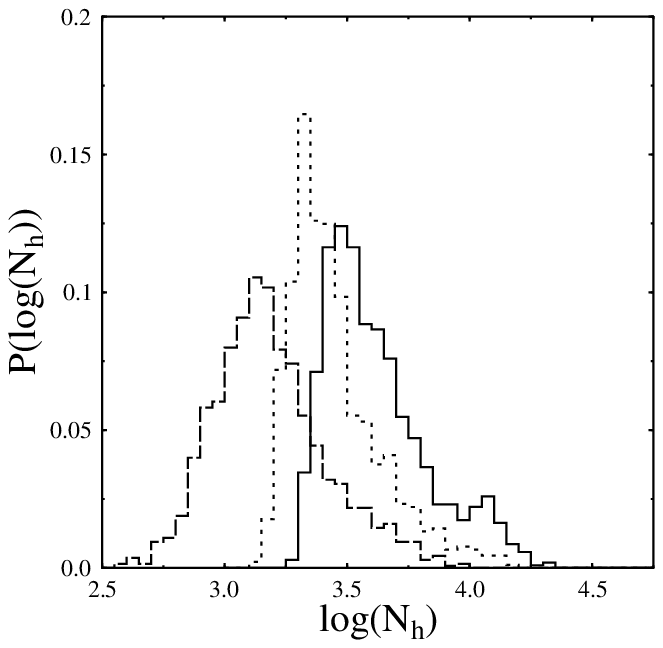}
\caption{Multiplicity distribution of hadrons in EAS with size
$N_e=10^6 \div 10^7$ detected at Chacaltaya and initiated by primary
protons (dashed), iron nuclei (dotted) and strangelets with $A_0=400$
(solid histogram).}
\end{minipage}\hspace{2mm}
\begin{minipage}[t]{0.475\linewidth}
\centering
\includegraphics[height=7.cm,width=7.cm]{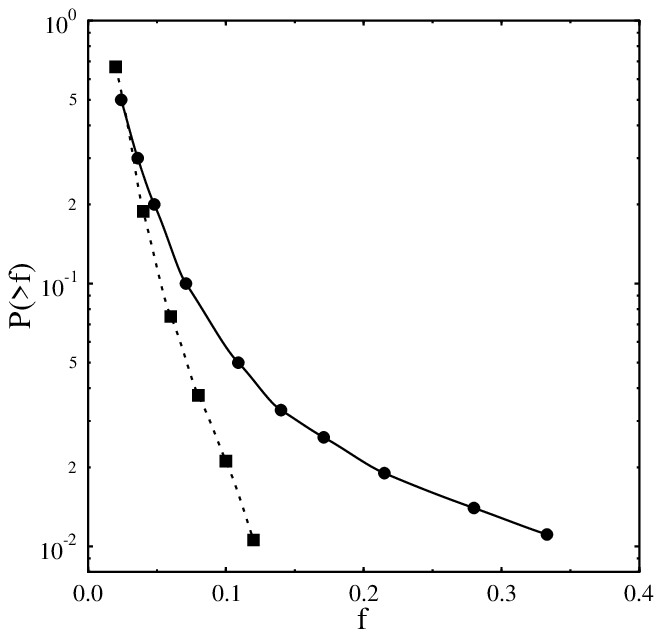}
\caption{Integral $f=E/ \sum E$ distribution in families with $\sum
E = 100 \div 200$ TeV initiated by primary protons (dots) and SQM
(squares).}
\end{minipage}
\end{figure}
The characteristic features in strangelets propagation in the
atmosphere are illustrated  in Figs. 1 and 2. Many characteristics of
the families and EAS are sensitive to existence of primary
strangelets. As an illustration we show in Fig. 5 our expectations
for the corresponding distributions of hadrons in EAS detected at
Chacaltaya. Analysis of gamma-ray families induced by cosmic rays
also shows significant differences between SQM and "normal" hadronic
matter. On Fig. 6 we show differences in $f=E/ \sum E$ distribution
in families with $\sum E = 100 \div 200$ TeV initiated by primary
protons and SQM. Families from SQM are characterized by soft energy
spectrum (being also much more broader) then those induced by primary
proton. 
The examples of results obtained for muons in EAS with and without
strangelets are presented in Figs. 7 and 8.

\section{Muon bundles from CosmoLEP}

We would like to bring ones attention to the data from the cosmic-ray
run of the ALEPH detector at the CosmoLEP experiment. Data archives
from the ALEPH runs have revealed a substantial collection of cosmic
ray muon events \cite{Cosmo}. More than $3.7 \cdot 10^5$ muon events
have been recorded in the effective run time $10^6$ seconds.
Multi-muon events observed in the $16$ $m^2$ time projection chamber
with momentum cut-off $70$ $GeV$ have been analyzed and good
agreement with the Monte Carlo simulations obtained for
multiplicities $N_{\mu}$ between 2 and 40. However there are 5 events
witch unexpectedly large multiplicities $N_{\mu}$ (up to 150) which
cannot be explained, even assuming pure iron primaries.
\vspace{-1cm}
\begin{figure}[h]
\begin{minipage}[t]{0.475\linewidth}
\centering
\includegraphics[height=7.cm,width=7.cm]{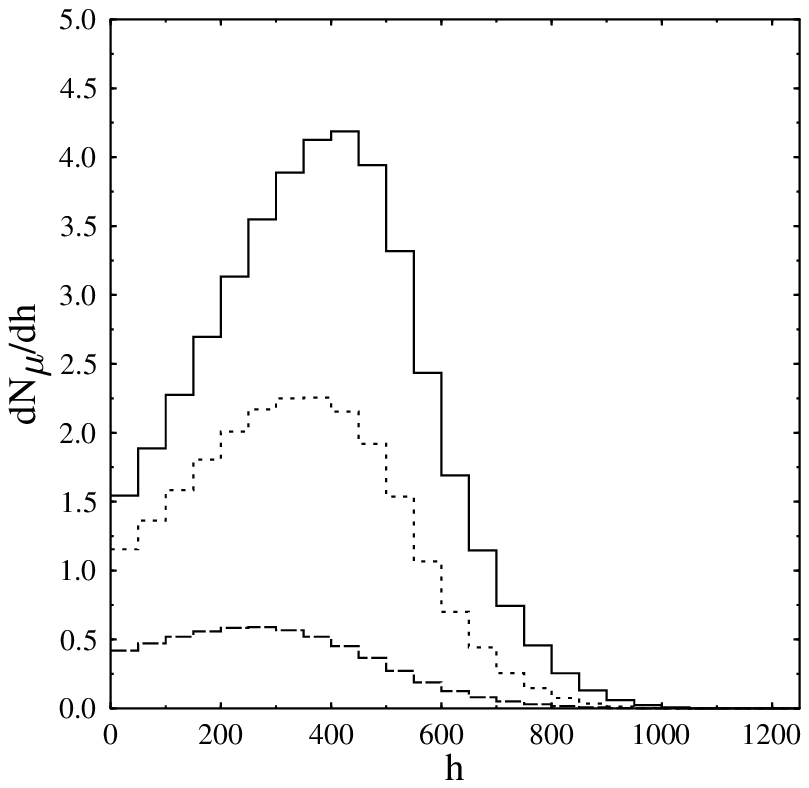}
\caption{Depth of origin of muons in EAS originated by primary protons
(dashed), iron nuclei (dotted) and strangelet with $A_0=400$ (solid).}
\end{minipage}\hspace{2mm}
\begin{minipage}[t]{0.475\linewidth}
\centering
\includegraphics[height=7.cm,width=7.cm]{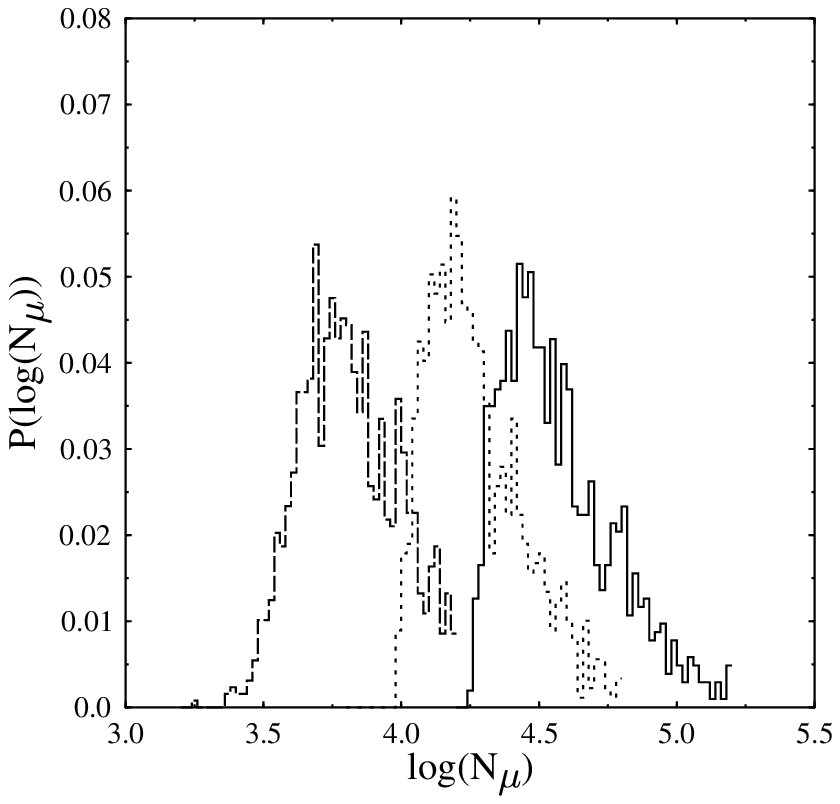}
\caption{Multiplicity distributions of muons in EAS with size $N_e=
10^6 - 10^7$ detected at Chacaltaya and indicated by primary
protons (dashed), iron nuclei (dotted) and strangelets
with $A_0=400$ (solid).}
\end{minipage}
\end{figure}

We shall estimate the production of muon bundles of extremely high
multiplicity in collisions of strangelets with atmospheric nuclei \cite{Ryb}.
Monte Carlo simulation describes the interaction of the primary
particles at the top of atmosphere and follows the resulting
electromagnetic and hadronic cascades through the atmosphere down to
the observation level. The integral multiplicity distribution of muons
from ALEPH data is compared with our simulations in Fig. 9. We have
used here the so-called "normal" chemical composition of primaries with
40\% protons, 20\% helium, 20\% CNO mixture, 10\% Ne-S mixture, and
10\% Fe. It can describe low multiplicity ($N_{\mu} \leq 20$) region
only. Muon multiplicity from strangelet induced showers are very broad.
As can be seen, the small amount of strangelets (with the smallest
possible mass number $A=400$ (the critical mass to be $A_{crit}=320$
here)) in the primary flux can accommodate experimental data. Taking
into account the registration efficiency for different types of
primaries one can estimate the amount of strangelets in the primary
cosmic flux. In order to describe the observed rate of high
multiplicity events one needs the relative flux of strangelets
$F_S/F_{tot} \simeq 2.4 \cdot 10^{-5}$ (at the same energy per
particle).

\vspace{-1cm}
\begin{figure}[h]
\begin{minipage}[t]{0.475\linewidth}
\centering
\includegraphics[height=7.cm,width=7.cm]{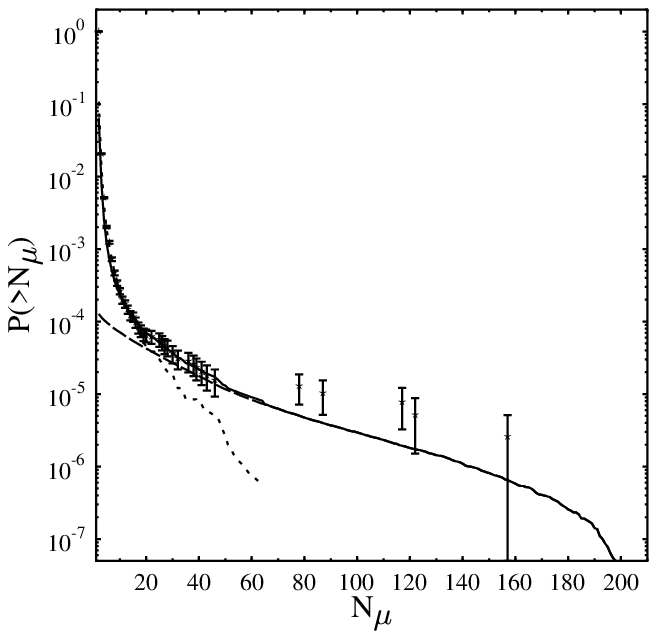}
\caption{Integral multiplicity distribution of muons for the
CosmoLEP data \cite{Ryb} (stars). Monte Carlo simulations for primary nuclei
with "normal" composition (dotted line) and for primary strangelets
with $A=400$ (broken line). Full line shows the summary (calculated)
distribution.}
\end{minipage}\hspace{2mm}
\begin{minipage}[t]{0.475\linewidth}
\centering
\includegraphics[height=7.cm,width=7.cm]{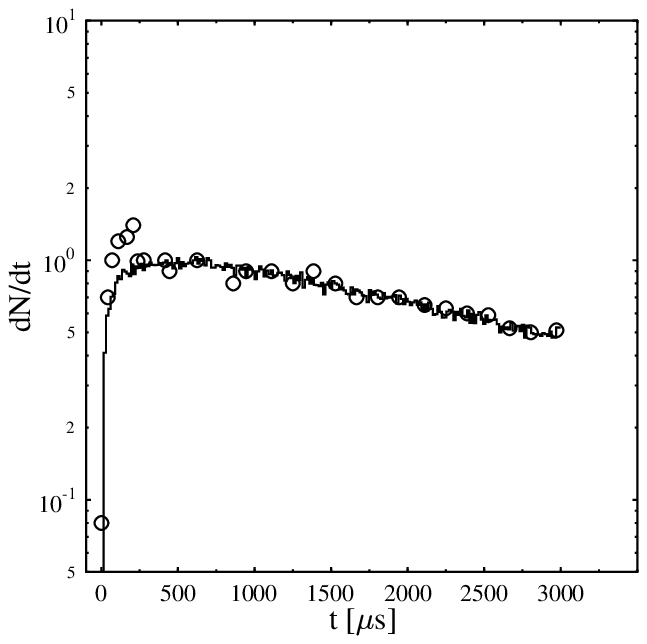}
\caption{Temporal distribution of highest multiplicity neutron
event (circles) recorded by boron neutron counter \cite{Anton} compared with
Monte Carlo simulations (full line) with mean lifetime of strangelet
$\tau=2000$ $\mu s$.}
\end{minipage}
\end{figure}
It can be interesting to point out that the high multiplicity events
discussed here (with $N_{\mu} \simeq 110$ recorded on $16$ $m^{2}$)
corresponds to $\sim 5600$ muons with $E_{\mu} \geq 70$ $GeV$ (or
$1000$ muons with energy above $220$ $GeV$). These numbers are in
surprisingly good agreement with results from other experiments like
Baksan Valley where 7 events with more than $3000$ muons of energy
$220$ $GeV$ were observed \cite{Baka}.

\section{Delayed neutrons}

In the last years some evidences have been found \cite{Aus} for the existence
of abnormal large events in neutron monitors that we shall call delayed
neutrons. These phenomena could not be explained by the known
mechanisms of hadronic cascades development. 

On the other hand such delayed neutrons may appear after decay of
small, unstable strangelet, which was created as a result of
interaction of primary cosmic rays with air nuclei. Mean lifetime of
that strangelet may be few thousands $\mu s$ long \cite{Berg,Craw}.
We calculated arrival time distribution of neutrons, like shown in 
\cite{Amb}. We used this distribution for simulation of the time
distribution of delayed neutrons which appear after decay of the
strangelet. The energy spectrum of evaporated neutrons followed the
Planck distribution with $T=4$ MeV. The integration over energy was 
performed in the energy interval $1 - 50$ MeV. In Fig. 10 we show
experimental data from standard neutron supermonitor 18NM64
\cite{Anton} in comparison with our simulations. If we assume mean
lifetime of a strangelet being $\tau \simeq 2000$ $\mu s$ it
describes satisfactory experimental data.

\section{Interactions with background radiation}

In analysis of cosmic rays propagation through the Universe one should
consider their interaction with background radiation. In the region of
high energies, the main processes are: pair production $p + \gamma
\rightarrow p + e^{+}e^{-}$, photoproduction $p + \gamma \rightarrow p
+ \pi^{0}$ and nuclei photodisintegration. The energy losses of nuclei
with mass $A$ during their interaction with background photons given by
\cite{Berez} shows remarkable $Z^{2}/A$ behaviour ($\sim A^{-1/3}$ in
the case of strangelets). In Fig. 11 we show the energy losses of
nuclei and strangelets arised due to interactions with background
radiation. As can be seen, because of their large mass number $A$,
strangelets can propagate through the Universe with very small energy
losses. Critical energy for strangelets is seemighty larger then this
for protons (c.f. Fig. 12). 
This can lead to the idea that extremely energetic cosmic rays are
not the result of the acceleration of protons, but rather of the decay
of unstable primordial objects. We expect that strangelets may
be highly energetic primordial remnants of the Big Bang falling into
this category.

\section{Abundance of strangelets}

We can estimate flux of strangelets on the border of atmosphere
using the existing experimental data taken at different atmospheric 
depths \cite{WWh}. Interpreting them in terms of our scenarios of
propagation of strangelets in the atmosphere, the flux of strangelets
hitting the Earths atmosphere as a function of mass is evaluated. The
results can be parametrized as follows: $\propto A_{0}^{-7.5}$. The
estimated flux of strangelets is consistent (c.f. Fig. 13) with the
predicted astrophysical limits and the upper limits given
experimentally \cite{Price}. 

\vspace{-1cm}
\begin{figure}[h]
\begin{minipage}[t]{0.475\linewidth}
\centering
\includegraphics[height=7.cm,width=7.cm]{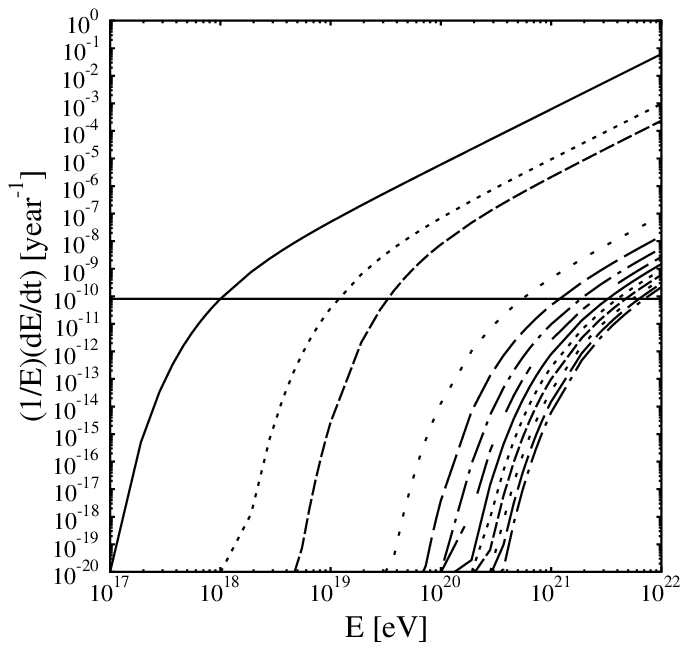}
\caption{Energy losses of protons, O, Fe, and some strangelets with
mass number $A=320,640,...,3200$ (respectively from left) due to
interaction with cosmic background radiation. Horizontal line shows red
shift limit.}
\end{minipage}\hspace{2mm}
\begin{minipage}[t]{0.475\linewidth}
\centering
\includegraphics[height=7.cm,width=7.cm]{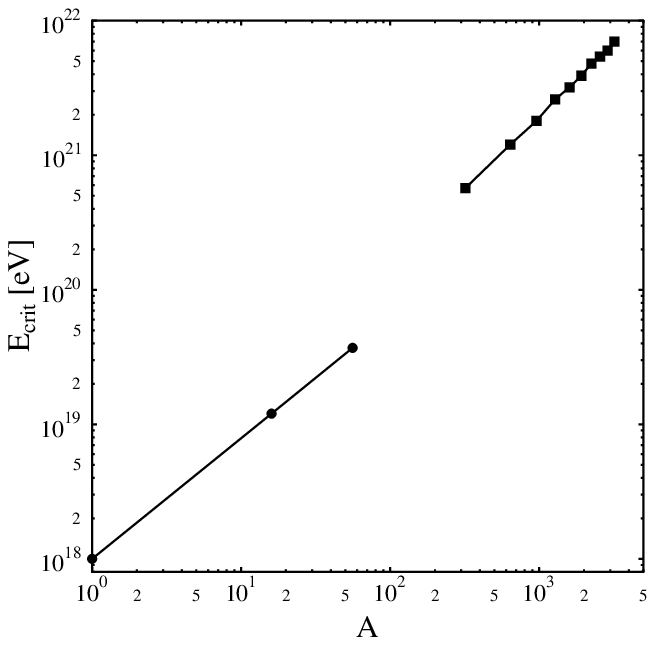}
\caption{GZK cutoff $E_{crit}$ as a function of mass number $A$.
Dots represents normal nuclei, squares - strangelets.}
\end{minipage}
\end{figure}

It is interesting to note that essentially the same power behaviour is
observed also for occurrence of normal nuclei in the Universe. Namely,
combination of existing data \cite{Zhdanov} on the chemical composition (for
normal nuclei, i.e., $Z < 100$ or $A < 250$) comprising CR (at
relativistic energies) and different astrophysical objects (like solar
system matter, Earth's core or Sun's atmosphere) can be described by
the formula $N(A) \propto A^{-7.5}$ (c.f. Fig. 14).

\section{Summary and conclusions}
Problem of possible Strange Quark Matter existence in the Universe were
discussed from many year. There are large amount of phenomena that
cannot be explain without assuming of existence of SQM. We demonstrated
only few examples from various fields of cosmic ray physics. In
particular we showed that extremely high energy cosmic rays,
exceeding the usual GZK limit, can indeed exists provided they are
composed of strangelets. Strangelets can be therefore considered as
natural candidates for looked for unstable primordial objects
(remnant of the Big Bang), decay of which can provide such energetic 
cosmic rays.

\vspace{-1cm}
\begin{figure}[h]
\begin{minipage}[t]{0.475\linewidth}
\centering
\includegraphics[height=7.cm,width=7.cm]{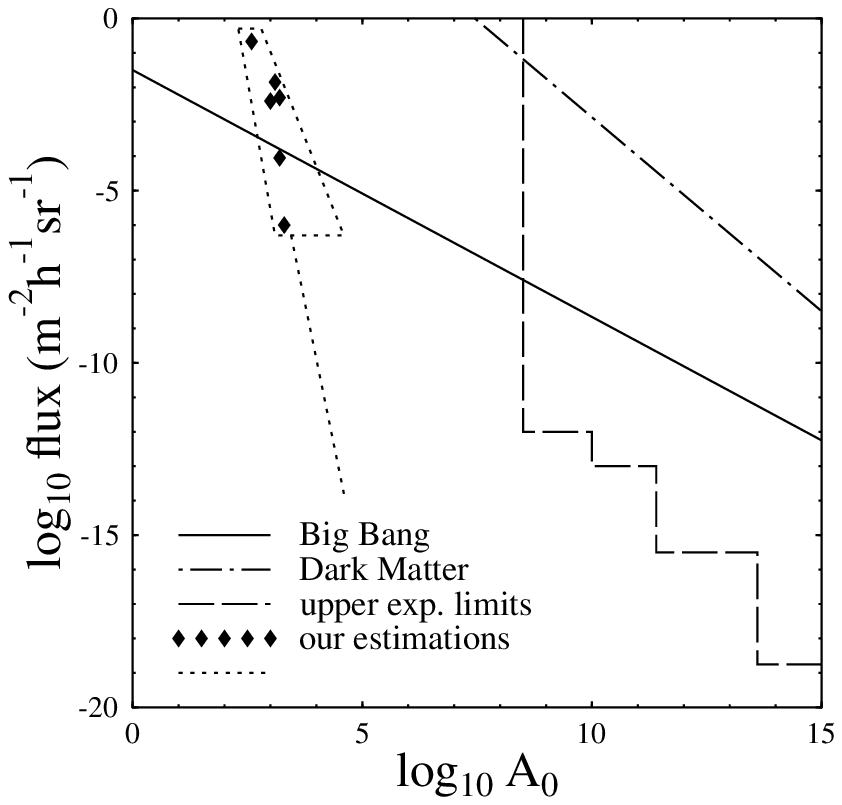}
\caption{The expected flux (our results) of strangelets compared
with the upper experimental limits, compiled by Price \cite{Zhdanov}, and
predicted astrophysical limits: Big Bang estimation comes from
nucleosynthesis with quark nuggets formation; Dark Matter one comes
from local flux assuming that galactic halo density is given solely by
quark nuggets.}
\end{minipage}\hspace{2mm}
\begin{minipage}[t]{0.475\linewidth}
\centering
\includegraphics[height=7.cm,width=7.cm]{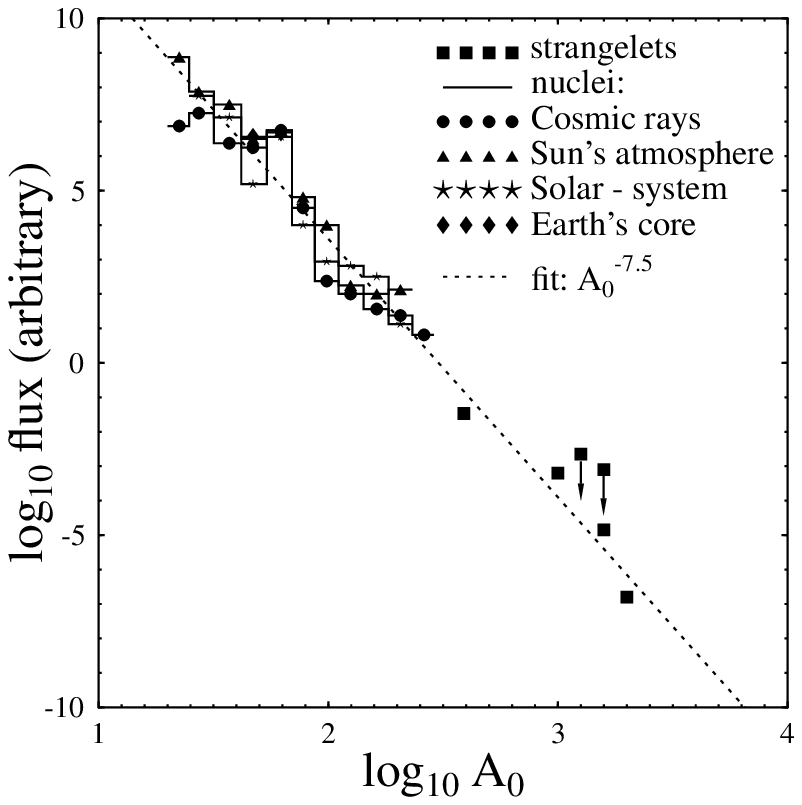}
\caption{Comparison of the estimated mass spectrum $N(A_{0})$ for
strangelets with the known abundance of elements in the Universe \cite{Grei}.}
\end{minipage}
\end{figure}

Let us finally mention a few words on problems of observability of
strangelets in heavy-ion collisions. One must stress from the very
beginning that those are different objects from the SQM originated
astrophysically in what concerns their size. In comparison with the
former they will be small, fastly decaying (because of the dominance
of the surface effects) object. Nevertheless, such strangelets 
can be to originate in the hadronization  of a QGP  fireball of any
high-baryon density and baryochemical potential, which is formed in
ultrarelavistic nucleus-nucleus collisions. Metastable droplets of
strange quark matter can be formed during the phase transition. This is
due to the production of kaons (containing $\overline{s}$ quarks) in
the hadron phase while $s$ quarks remain in the deconfined phase. The
$s-\overline{s}$ separation results in a strong enhancement of the
$s$-quark abundance in the quark phase \cite{Grei}. Several
experimental searches for SQM had been carried out at the AGS at the
BNL and at the SPS at the CERN. 
\begin{itemize}
\item The experiment E864 (BNL-AGS) aims to study heavy-ion
collisions in an open geometry spectrometer in $11.5 A GeV$ $Au+Pb$
collisions \cite{Arm}. So far no neutral strangelets was found (the upper
limit of strangelets obtained is of the order of $10^{-8}$ per
central collisions).
\item NA52 is the only dedicated experiment at CERN 
which searches for strangelets in sulfur-tungsten and lead-lead
interactions. Statistics of several $10^{12}$ interactions have been
collected. This data set addresses the question of a short-lived
strangelet candidate with mass $7.4$ $GeV$ and charge $Z=-1$ \cite{Kling}.
\end{itemize}
A new experimental approach proposed recently is to search for
strangelets by using unconventional strangelet signature, namely by
looking at the energy deposition pattern (c.f. 5.1) trying to observe
strangelets occuring as deeply penetrating objects. The corresponding
detector system CASTOR is proposed as an integral part of the planned
CERN-LHC ALICE experiment \cite{Angelis}. CASTOR will cover the very forward
rapidity region ($5.6 < \eta < 7.2$). This is assumed to be a
preferred region to create a dense quark matter fireball because it
is baryon-rich. This could probe a strangelets separation mechanism
like distillation. \\

The partial support of Polish Committee for Scientific Research
(grants 2P03B 011 18, 5P03B 091 21 and 621/ E-78/ SPUB/
CERN/P-03/DZ4/99) is acknowledged.


\end{document}